\documentclass{moriond}

\def\be{\begin{equation}}
\def\ee{\end{equation}}
\def\bea{\begin{eqnarray}}
\def\eea{\end{eqnarray}}

\newcommand{\met}{\ensuremath{E_{T}^{miss}}}
\newcommand{\tauh}{\ensuremath{\tau_{\mathrm h}}}
\newcommand{\ifb}{\ensuremath{\mathrm{fb}^{-1}}}
\newcommand{\chiz}{\ensuremath{\widetilde{\chi}^{0}}}
\newcommand{\chipm}{\ensuremath{\widetilde{\chi}^{\pm}}}
\newcommand{\chip}{\ensuremath{\widetilde{\chi}^{+}}}
\newcommand{\chim}{\ensuremath{\widetilde{\chi}^{-}}}
\newcommand{\chitn}{\ensuremath{\chiz_{2}}}
\newcommand{\lsp}{\ensuremath{\chiz_{1}}}
\newcommand{\chipmo}{\ensuremath{\chipm_{1}}}
\newcommand{\chipo}{\ensuremath{\chip_{1}}}
\newcommand{\chimo}{\ensuremath{\chim_{1}}}
\newcommand{\slep}{\ensuremath{\widetilde{\ell}}}
\newcommand{\grav}{\ensuremath{\widetilde{G}}}

\newcommand{\Photo}{}

\begin{document}
\vspace*{4cm}
\title{SEARCHES FOR WEAKLY INTERACTING SUPERSYMMETRIC PARTICLES}

\author{ D. OLIVITO }

\address{Department of Physics, University of California, San Diego \\
9500 Gilman Dr, La Jolla, CA 92093, USA}

\maketitle\abstracts{
The ATLAS and CMS collaborations have performed several targeted searches for weakly interacting supersymmetric particles, using data collected in proton-proton collisions at sqrt(s)=8~TeV at the LHC.  No significant deviations from standard model predictions have been observed.  These proceedings summarize some of the latest experimental results and their interpretations.
}

\section{Introduction}

Discovering or constraining supersymmetry (SUSY) is one of the main experimental goals
of the ATLAS~\cite{atlas_det} and CMS~\cite{cms_det} collaborations at the Large Hadron Collider (LHC).  The weakly interacting SUSY
particles, charginos (\chipm), neutralinos (\chiz), and sleptons (\slep), could be pair-produced 
through electroweak processes in $pp$ collisions at the LHC.   
Depending on the sparticle spectrum and the composition of the chargino and neutralino mass eigenstates,
different production and decay modes would be available, leading to different final states experimentally.

The ATLAS and CMS collaborations have performed a broad array of searches to target the various possible final
states, and these proceedings summarize some of the latest results as of the Moriond QCD 2014 conference.  
As the results below highlight, one of the novel approaches is using the recently discovered Higgs boson
to search for physics beyond the standard model (SM).  
All the searches described here use the full dataset collected at $\sqrt{s}$=8~TeV, corresponding 
to an integrated luminosity of about 20~\ifb.  As they assume R-parity conservation, they all require missing 
transverse energy (\met) consistent with the presence of undetected LSPs.

\section{Search in the Three Lepton + \met\ Final State}
\label{sec:atlas_3lep}

ATLAS performed a search in the three lepton + \met\ final state~\cite{atlas_3lep}.  It is sensitive to chargino-neutralino
production in several decay modes, such as:
\begin{itemize}
\item $\chipmo\chitn \rightarrow \slep\nu\slep\ell \rightarrow \ell\nu\lsp\ell\ell\lsp$
\item $\chipmo\chitn \rightarrow W\lsp Z\lsp \rightarrow \ell\nu\lsp\ell\ell\lsp$
\item $\chipmo\chitn \rightarrow W\lsp H\lsp \rightarrow \ell\nu\lsp\ell\ell\lsp+X$
\end{itemize}
where $\ell = e,\mu,\tau$.  In the first case, the sleptons are lighter than the chargino and neutralino, 
thus accessible in decays, while in the later cases they are assumed to be heavy and decoupled. 
In the last case, the Higgs boson decays via $WW$, $ZZ$, or $\tau\tau$ to produce at least two leptons.

The search selects events with exactly three leptons, using $e$, $\mu$, and up to two hadronically-decaying 
$\tau$ candidates (\tauh), and events are binned according to their lepton content, with
further selections applied to enhance sensitivity to each of the final states given above.  For example, signal regions with 
either one or two \tauh\ candidates target the final state of $W\lsp H\lsp \rightarrow \ell\nu\tau\tau + \met$ 
by requiring the visible dilepton mass to be consistent with the Higgs boson mass.
In total the analysis defines 24 signal regions, 
with all mutually exclusive except the two bins requiring two \tauh\ candidates.  

The primary SM backgrounds come from processes producing three or more prompt leptons, plus reducible sources where at least
one lepton is non-prompt or a misreconstructed jet.  The former is predicted from simulation and the latter from control
regions in data.  
No significant deviations from the SM predictions are observed.  The results from interpreting this search
in the context of simplified models are shown in Fig.~\ref{fig:summary}~(left) with the curves labeled ``3l.''  Assuming a massless 
LSP, chargino-neutralino production is probed up to a common chargino-neutralino mass of about 300--700~GeV, depending
on the decay mode.

Similar searches have also been performed by CMS~\cite{cms_multilep,cms_ewkino,cms_whmet}. 
The CMS searches do not consider events with two \tauh\ leptons, but Ref.~\cite{cms_multilep} in particular 
has a more inclusive selection with bins in e.g. $H_{T}$ (the scalar sum of jet $p_{T}$) and the presence of absence of a b jet.
No significant deviations are seen in these searches.  The results are combined with other search channels and
appear in Fig.~\ref{fig:summary}~(right).

\begin{figure}
\begin{center}
\includegraphics[width=0.44\linewidth]{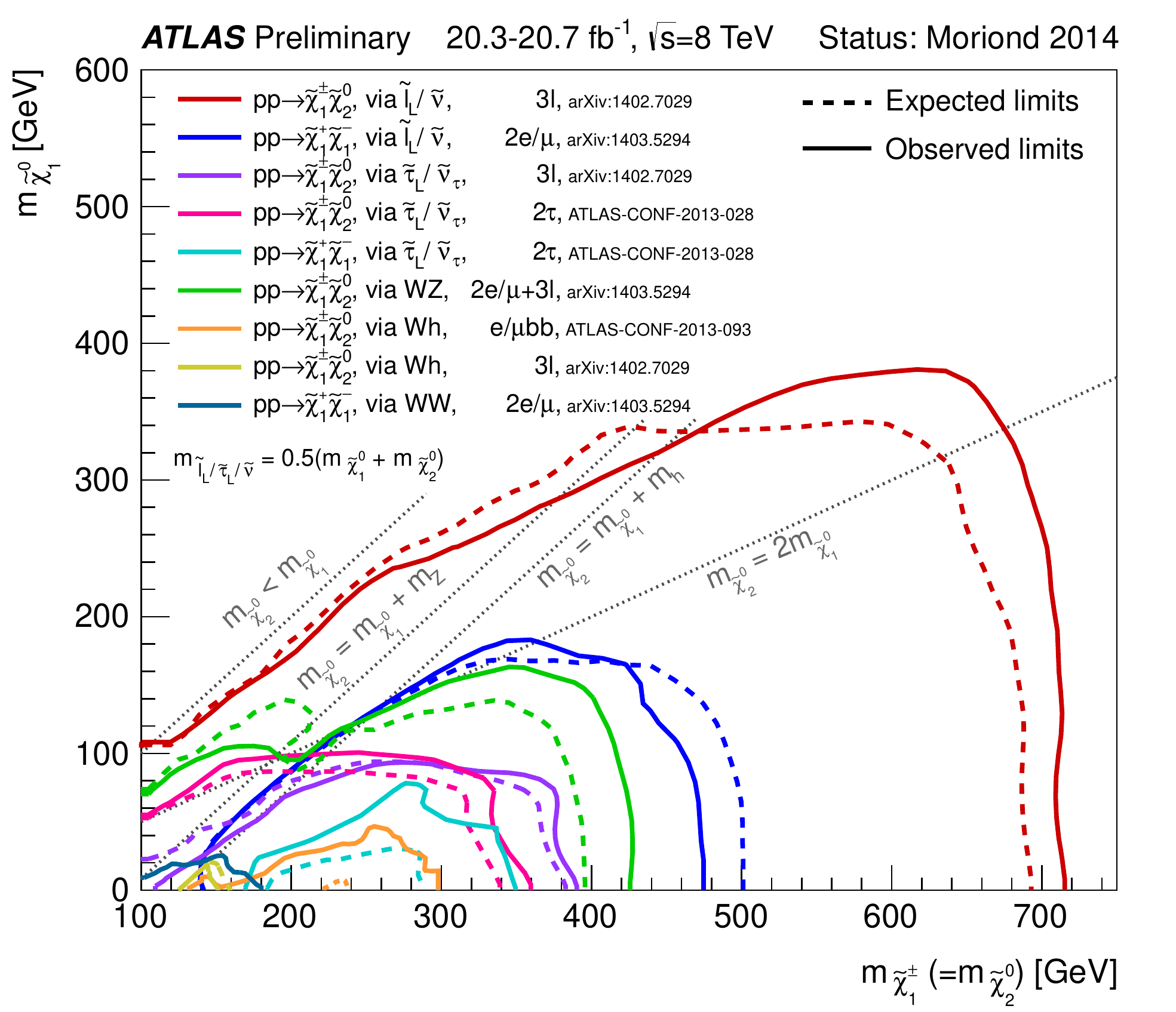}
\includegraphics[width=0.5\linewidth]{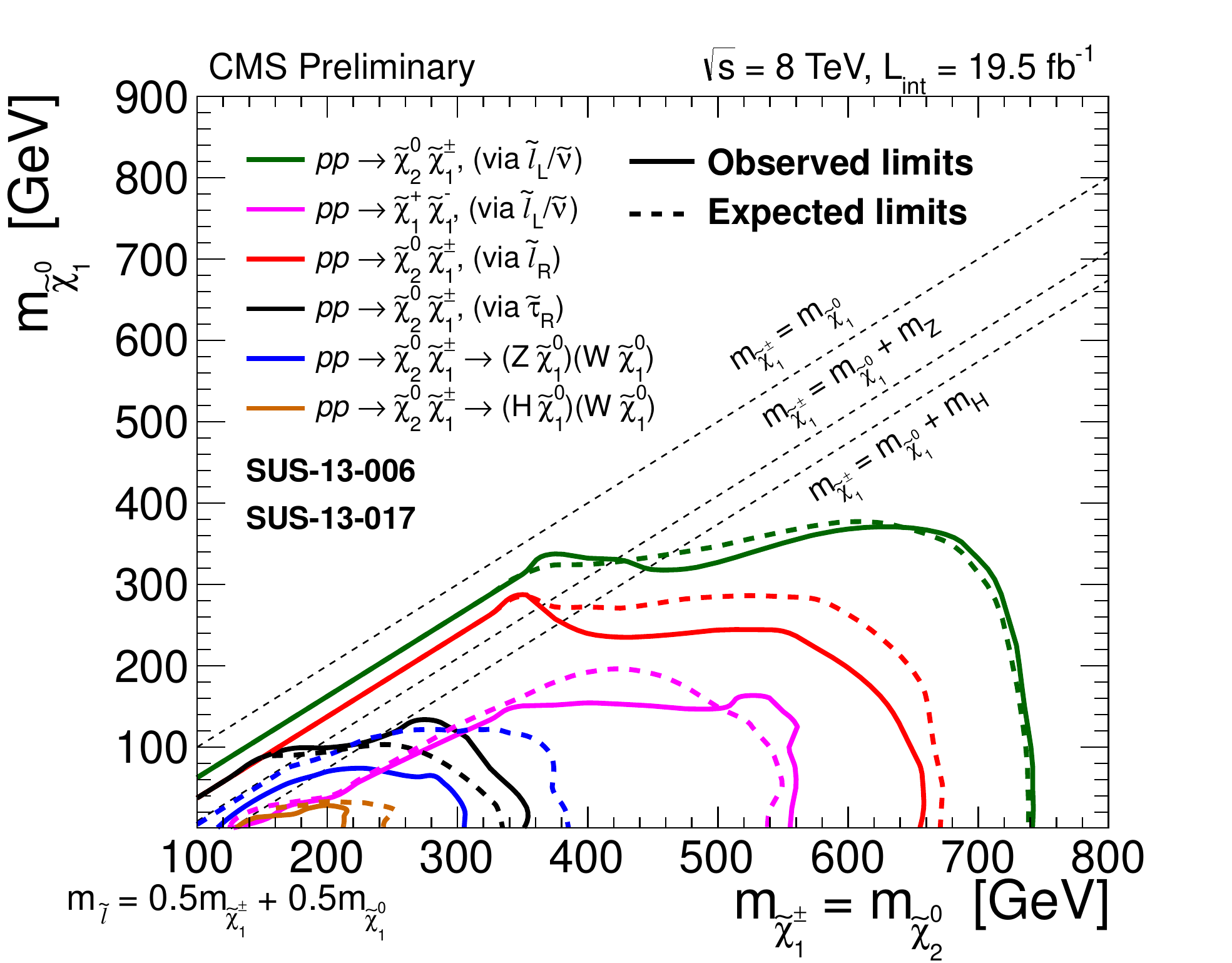}
\caption[]{Summary of simplified model interpretations for ATLAS~(left) and CMS~(right) searches for pair production
of charginos and neutralinos, for different production and decay modes.}
\label{fig:summary}
\end{center}
\end{figure}

\section{Search in the Two Lepton + \met\ Final State}

ATLAS performed a search in the two lepton + \met\ final state~\cite{atlas_2lep} with sensitivity to processes such as:
\begin{itemize}
\item $\chipo\chimo \rightarrow W^{+}\lsp W^{-}\lsp \rightarrow \ell^{+}\nu\lsp\ell^{-}\nu\lsp$ 
or $\chipo\chimo \rightarrow \slep^{+}\nu\slep^{-}\nu \rightarrow \ell^{+}\nu\lsp\ell^{-}\nu\lsp$
\item $\chipmo\chitn \rightarrow W\lsp Z\lsp \rightarrow qq'\lsp\ell^{+}\ell^{-}\lsp$
\item $\slep^{+}\slep^{-} \rightarrow \ell^{+}\lsp\ell^{-}\lsp$
\end{itemize}
Events with two opposite-sign leptons ($e$ or $\mu$) are selected and binned according to their
lepton content, including the presence or absence of a $Z$ boson candidate based on invariant mass.
Seven signal regions are defined with additional kinematic selections to target the final states above.
In events with a $Z$ boson candidate, the largest backgrounds come from $Z$+jets and $t\bar{t}$, while
in events without a $Z$ candidate, the largest backgrounds are from $WW$ and $t\bar{t}$.  The main backgrounds
are either predicted from data or taken from simulation and normalized in data control regions.
No significant deviations are observed.  

Interpretations of the results are shown in Fig.~\ref{fig:summary}~(left) with the curves labeled ``2$e/\mu$.''
The first constraints from the LHC on the process $\chipo\chimo \rightarrow W^{+}\lsp W^{-}\lsp$ are derived, probing chargino
masses between 100 and 160~GeV for a massless LSP.  CMS has performed similar searches~\cite{cms_ewkino}, with the main
difference being no targeted search for the process $\chipo\chimo \rightarrow W^{+}\lsp W^{-}\lsp$.  The results are included
in the interpretations of Fig.~\ref{fig:summary}~(right).

\section{Search in the WH + \met\ Final State}

CMS performed a search for the process $\chipmo\chitn \rightarrow W\lsp H\lsp$ using events with one, 
two, or at least three leptons to target different decays of the Higgs boson~\cite{cms_whmet}.  The one 
lepton search targets the final state $W\lsp H\lsp \rightarrow \ell\nu b\bar{b} + \met$ and gives the best sensitivity 
in most of the available parameter space.  It requires exactly one lepton ($e$,$\mu$) and two b jets consistent
with the Higgs boson mass, plus additional kinematic requirements to reduce the large backgrounds from $t\bar{t}$,
$W$+jets, and $WZ$ production.  The backgrounds are mainly predicted from simulation, with corrections from 
orthogonal control regions in data.

The two lepton search looks for the final state 
$W^{\pm}\lsp H\lsp \rightarrow W^{\pm}\lsp W^{\pm}W^{\mp}\lsp \rightarrow \ell^{\pm}\nu \ell^{\pm}\nu jj  + \met$
by requiring exactly two same-sign leptons and two or three jets.  Additional kinematic selections are applied to
reject backgrounds from rare SM processes, events with misidentified leptons, and events with the electron charge
misidentified.  The background from misidentified leptons is predicted from data, while the others are predicted 
from simulation.

No significant deviations are seen in any of the channels, and the results are combined to place the limits shown
in Fig.~\ref{fig:summary}~(right).  Assuming the chargino and neutralino are degenerate in mass and that the LSP is
massless, the results probe up to about 200~GeV in the mass of the chargino.

In addition to the three lepton search described in Sec.~\ref{sec:atlas_3lep}, ATLAS performed a similar 
search to CMS in the one lepton channel~\cite{atlas_1lep}.  No deviations were observed, and the results 
are shown in Fig.~\ref{fig:summary}~(left) with the curve labeled ``$e/\mu$bb.''

\section{Search in the Four b jet + \met\ Final State}

CMS performed a search in the $HH+\met \rightarrow (b\bar{b})(b\bar{b})+\met$ final state~\cite{cms_4b}. 
Events are selected with exactly four or five jets and at least two b jets, with the most sensitive bins requiring
at least four b jets.  Jets are paired into two Higgs
boson candidates by minimizing the difference in invariant mass between the pairs.  The average mass of the 
two pairs then peaks at the Higgs boson mass for signal and is used to reject background, along with additional
kinematic requirements.  The main backgrounds arise from $t\bar{t}$ and QCD multijet production, and they are
predicted from orthogonal control regions in data.

No significant deviations from the SM predictions are observed, and limits are set in the context of a gauge-mediated
supersymmetry breaking (GMSB) model.
The massless gravitino is the LSP, and the higgsinos are assumed to be the next-lightest states and degenerate in mass.
The limit is shown in Fig.~\ref{fig:others}~(left).  None of the parameter space is excluded given the borderline expected 
sensitivity of the search and a slight upward fluctuation in data.

\begin{figure}
\begin{center}
\includegraphics[width=0.45\linewidth]{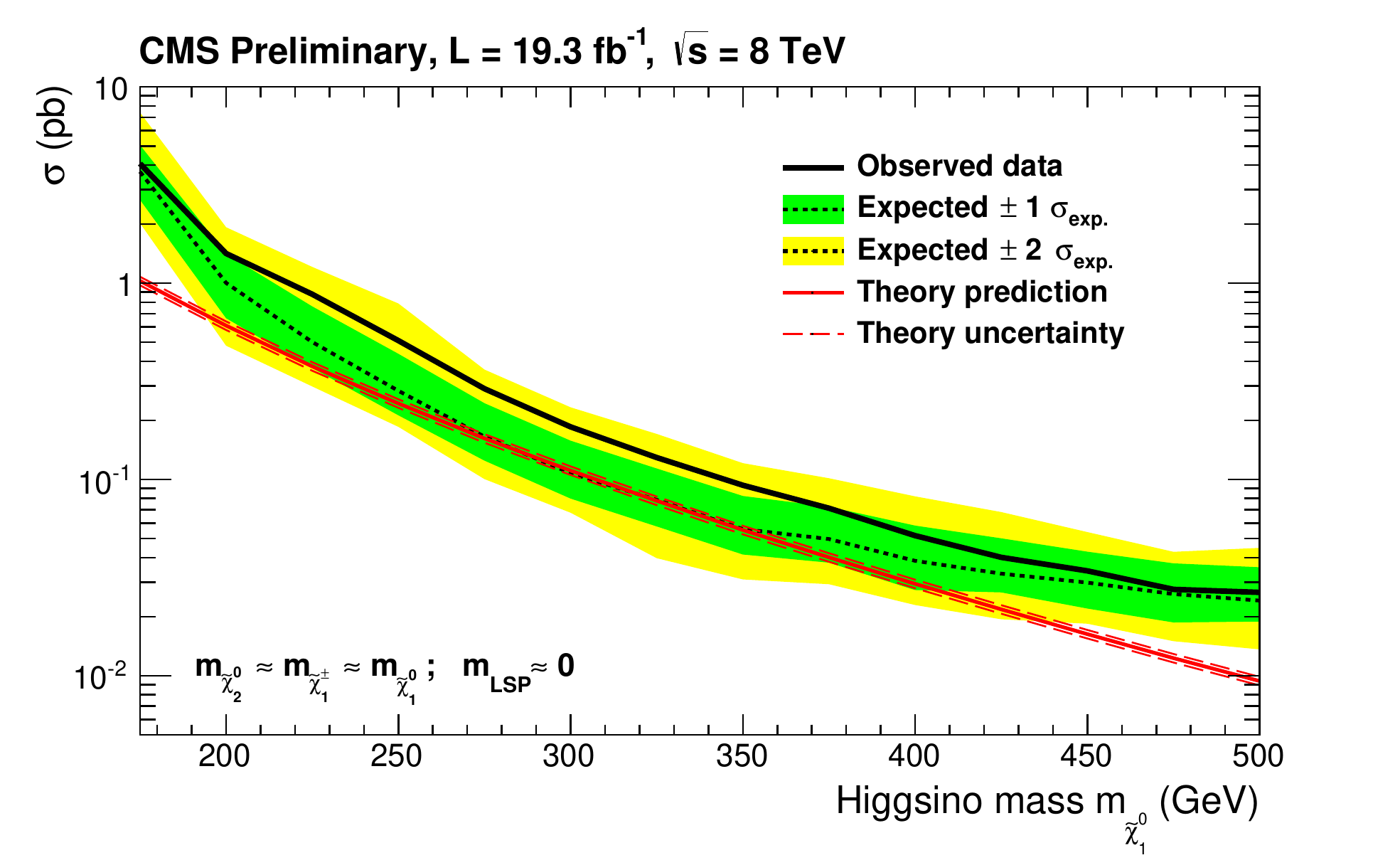}
\includegraphics[width=0.45\linewidth]{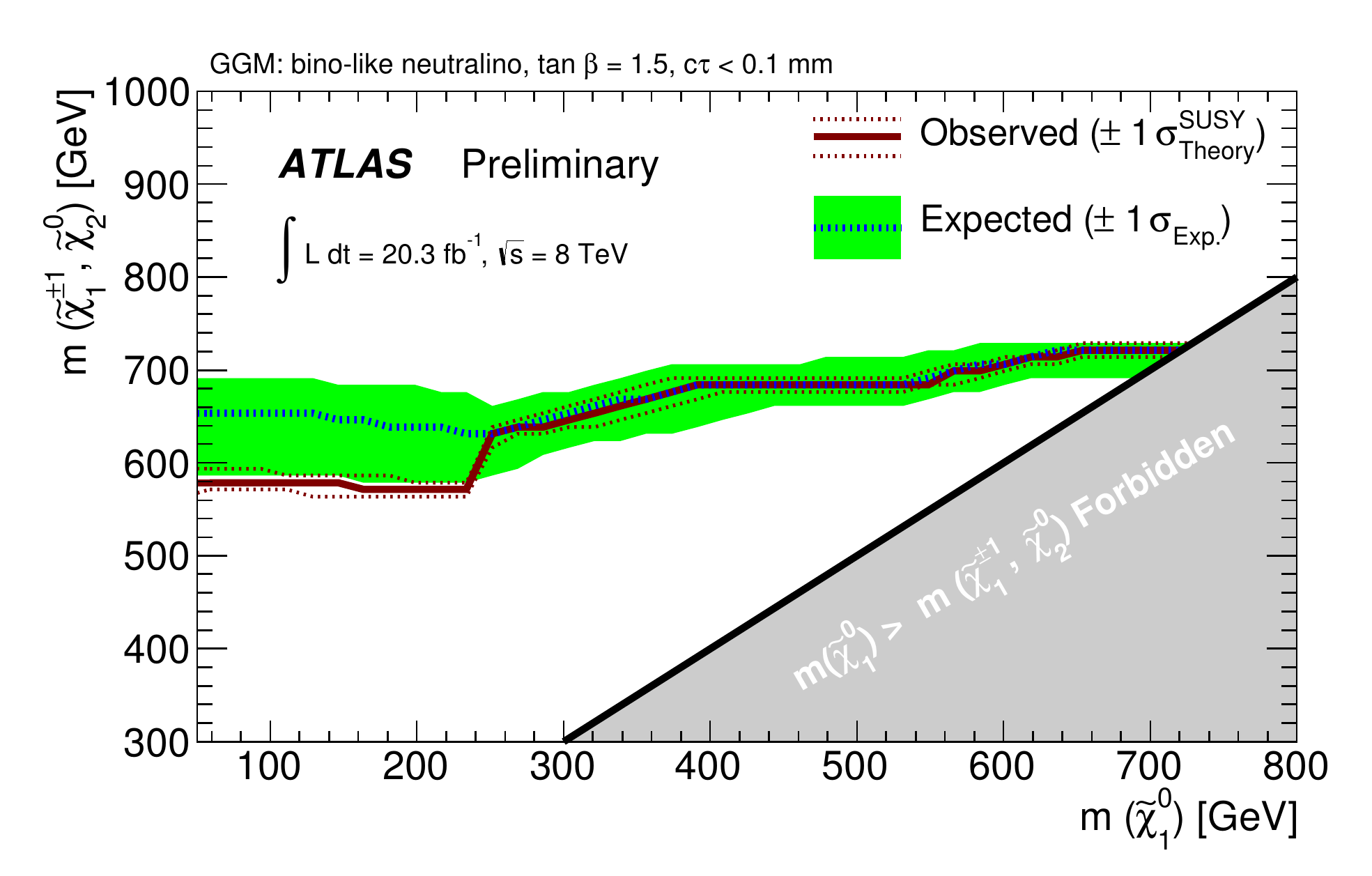}
\caption[]{(left) Limit from the CMS four b jet + \met\ analysis interpreted in a GMSB model. 
(right) Limit from the ATLAS two photon + \met\ analysis interpreted for weak production in a GGM model.}
\label{fig:others}
\end{center}
\end{figure}

\section{Search in the Two Photon + \met\ Final State}

ATLAS performed a search in the two photon + \met\ final state~\cite{atlas_2gamma}.  It targets both strong and weak
production in the context of a general gauge mediation (GGM) model by making additional requirements on $H_{T}$, 
defined as the scalar sum of the photons plus any jets or leptons.  Further kinematic requirements are also imposed.
The backgrounds come from multiple sources, including events with two prompt photons as well as events with one or two
jets or electrons misreconstructed as photons.  Most of the backgrounds are predicted from data using orthogonal control
regions. Five signal regions are defined, and no significant deviations from the SM predictions are observed.

The results are interpreted for weak production in a GGM model, with typical decay chains including:
\begin{itemize}
\item $\chipo\chimo \rightarrow W\lsp W\lsp \rightarrow \gamma\gamma+WW\grav\grav$
\item $\chipmo\chitn \rightarrow W\lsp Z\lsp \rightarrow \gamma\gamma+WZ\grav\grav$
\end{itemize} 
where \grav\ is the massless gravitino and the bosons decay either leptonically or hadronically.  
The limits are shown in Fig.~\ref{fig:others}~(right).

\section{Conclusions}

Several ATLAS and CMS searches for weakly interacting supersymmetric particles have been summarized here, 
with no significant deviations from SM predictions observed thus far.  Chargino-neutralino production is probed up to 
about 300--700~GeV in the chargino mass depending on the decay mode. Readers are encouraged to consult the documentation
for the individual analyses for more detail.

\end{document}



